\documentclass[%
 aip,
 amsmath,amssymb,
 reprint,%
]{revtex4-1}

\usepackage{graphicx}
\usepackage{dcolumn}
\usepackage{bm}

\usepackage[utf8]{inputenc}
\usepackage[T1]{fontenc}
\usepackage{mathptmx}
\usepackage{graphicx}
\usepackage{subeqnarray}
\usepackage{flushend}

\begin{document}

\preprint{AIP/123-QED}

\title[Black holes with a fluid of strings]{Black holes with a fluid of strings}

\author{J. M. Toledo}
\email{jefferson.m.toledo@gmail.com.}
 \altaffiliation{Departamento de F\'isica, Universidade Federal da Para\'iba, Caixa Postal 5008, CEP 58051-970, Jo\~ao Pessoa, PB, Brazil.}
\author{V. B. Bezerra}%
 \email{valdir@fisica.ufpb.br.}
\affiliation{ 
Departamento de F\'isica, Universidade Federal da Para\'iba, Caixa Postal 5008, CEP 58051-970, Jo\~ao Pessoa, PB, Brazil.
}%

\date{\today}

\begin{abstract}
We obtain an exact solution of Einstein's equations for a charged, static and spherically symmetric body, surrounded by a fluid of strings and with a cosmological constant. This corresponds to the Reissner-Nordstr\"om (de Sitter)-Anti de Sitter black hole surrounded by a fluid of strings. Some aspects concerning the horizons and geodesic motion are discussed. We also obtain the metric corresponding to the Kerr-Newman black hole surrounded by a fluid of strings, by using a method to construct the rotating solution from its counterpart static one. The horizons, ergoregions, and geodesic motion are analyzed. In both cases, the role played by the fluid of strings was pointed out. 
\end{abstract}

\maketitle

\section{Introduction}
\label{sec1}

In the early 1980s, Letelier \cite{letelier1981fluids} generalized the gauge-invariant model with a cloud of strings \cite{letelier1979clouds}, in the sense that pressure produced by the strings was now taken into account. In this context, he obtained the general solution of Einstien's field equations for a spherically symmetric body surrounded by a fluid of strings with this symmetry. Some years later, Soleng \cite{soleng1995dark} obtained an exact solution corresponding to the spacetime associated with a point mass surrounded by a static, spherically symmetric fluid of strings, assuming that this fluid is such that the transverse pressure is proportional to the energy density \cite{dymnikova1992vacuum, soleng1994correction}. Using this solution, Soleng \cite{soleng1995dark} concluded that when $|T^{t}_{\phantom{a}t}| \simeq |T^{t}_{\phantom{a}t}| \gg |T^{\Omega}_{\phantom{a}\Omega}|$, where $\Omega$ stands for both $\theta$ and $\phi$, the solution gives an $1/r$ correction to Newton gravitational force law, which means that it is possible, in principle, to explain the missing mass problem or, equivalently, the flat rotation curves of galaxies. Otherwise, in this non-relativistic scenario, in which an artificial set up is used, we cannot identify this fluid of strings as giving us a solution to the missing mass problem.

Taking into account this solution obtained by Soleng \cite{soleng1995dark}, we generalize it by considering an electromagnetic field as well as the cosmological constant. In this scenario, the solution describing the Reissner-Nordstr\"om-(de Sitter) Anti-de Sitter black hole surrounded by a fluid of strings is obtained. Additionally, we neglect the cosmological constant and constructed the metric corresponding to the Kerr-Newman black hole surrounded by a fluid of strings, using an algorithm with appropriate modification made recently \cite{azreg2014generating}. In both cases, we study the horizons and the geodesic motion, pointing out the role played by the fluid of strings.

This paper is organized as follows. In Sec. \ref{sec2}, we present a brief review concerning the solution obtained by Soleng \cite{soleng1995dark}. In Sec. \ref{sec3}, we obtain the solution of Einstein's equations corresponding to a static black hole with cosmological constant and surrounded by a fluid of strings. We also discuss, in this section, the horizons and geodesic motion. In Sec. \ref{sec4}, we obtain the Kerr-Newman black hole surrounded by a fluid of strings and discuss its horizons and geodesic motion of particles in this background. Finally, in Sec. \ref{sec5}, we present the concluding remarks.

\section{The static and uncharged black hole with a fluid of strings}
\label{sec2}

The world line of a moving particle with four-velocity given by $u^{\mu} = d x^\mu/ d \xi$, with $\xi$ being an independent parameter, can be described by $x = x(\xi)$. Otherwise, if we consider, instead of a particle, a moving infinitesimally thin string, thus, the trajectory corresponds to a two-dimensional worldsheet $\Sigma$, which can be obtained by \cite{letelier1979clouds}

\begin{equation} \label{eq1}
x^\mu = x^\mu (\xi^a), \qquad a=0,1,
\end{equation}

\noindent with $\xi^0$ and $\xi^1$ being timelike and spacelike parameters, respectively. Therefore, instead of the four-velocity, $u^{\mu}$ we have a bivector $\Sigma^{\mu \nu}$, such that \cite{letelier1979clouds}

\begin{equation} \label{eq2}
\Sigma^{\mu \nu} = \epsilon^{ab} \frac{\partial x^\mu}{\partial \xi^a }\frac{\partial x^\nu}{\partial \xi^b },
\end{equation}

\noindent where $\epsilon^{ab}$ is the two-dimensional Levi-Civita symbol, with $ \epsilon^{01}=- \epsilon^{10} = 1$.

It is worth emphasize that on this worldsheet, there will be an induced metric, $h_{ab}$, with $a, b = 0, 1$, such that,
\begin{equation} \label{eq2a}
h_{ab} = g_{\mu \nu} \frac{\partial x^\mu}{\partial \xi^a}\frac{\partial x^\nu}{\partial \xi^b},
\end{equation}
\noindent whose determinant we are indicatinh by $h$.

The energy-momentum tensor associated with a dust cloud is given by $T^{\mu \nu} = \rho u^\mu u^\nu$, with $u^\mu$ being the normalized four-velocity and $\rho$ being the proper density of the flow. Similarly, for a cloud of strings, we have \cite{letelier1979clouds}

\begin{equation} \label{eq3}
T^{\mu \nu}=\frac{\rho \Sigma^{\mu \beta} \Sigma^{\phantom{a}\nu}_\beta}{\sqrt{-h}},
\end{equation}

\noindent where $h=\frac{1}{2} \Sigma^{\mu \nu}\Sigma_{\mu \nu}$.

Now, if we take into account a perfect fluid with pressure $p$, it can be described by the stress-energy tensor $T^{\mu \nu} = (\rho+p) u^\mu u^\nu - p g^{\mu \nu}$. Similarly, considering a perfect fluid of strings with pressure $q$, we have the stress-energy tensor \cite{letelier1981fluids}

\begin{equation} \label{eq4}
T^{\mu \nu}=(q + \sqrt{-h} \rho)\frac{\Sigma^{\mu \beta} \Sigma^{\phantom{a}\nu}_\beta}{(-h)}+q g^{\mu \nu}.
\end{equation}

Considering the stress-energy tensor given by Eq. (\ref{eq4}), Soleng obtained the metric corresponding to a static black hole surrounded by a fluid of strings \cite{soleng1995dark}. To achieve that, the author considered that the metric corresponding to a static and spherically symmetric black hole can be written as

\begin{equation} \label{eq5}
ds^2 = - e^{2 \mu} dt^2 + e^{2 \lambda} dr^2+r^2 d\Omega^2,
\end{equation}

\noindent where $d\Omega^2 = d\theta^2 +r^2 d\phi^2$, and $\mu$ and $\lambda$ are functions of the radial coordinate only. Then, we assume that the components of the stress-energy tensor are related through the equations

\begin{subeqnarray} \label{eq6}
T^{t}_{\phantom{a}t} &=& T^{r}_{\phantom{a}r},\\
T^{t}_{\phantom{a}t} &=& - \alpha T^{\theta}_{\phantom{a}\theta} = - \alpha T^{\phi}_{\phantom{a}\phi},
\end{subeqnarray}

\noindent where $\alpha$ is a dimensionless constant. The energy-momentum tensor whose components are given by Eq. (\ref{eq6}), was interpreted as being associated with a kind of anisotropic fluid with spherical symmetry \cite{dymnikova1992vacuum, soleng1994correction}. Thus, writing Einstein's field equations in this context, Soleng \cite{soleng1995dark} obtained the following solution

\begin{equation} \label{eq7}
e^{2 \mu} = e^{-2 \lambda} = 1- \frac{2M}{r}+\left\{
\begin{array}{cc}
\epsilon l r^{-1} \log(\eta r) & \mbox{ for} \quad \alpha =2 \\
\epsilon \alpha( \alpha-2)^{-1} \left( \frac{l}{r}\right)^{2/\alpha} & \mbox{ for} \quad \alpha \neq 2,
\end{array}
\right.
\end{equation}

\noindent where $M$ is the black hole mass, $l$ and $\eta$ are positive integration constants and $\epsilon = \pm 1$ determines the sign of the energy density of the string fluid. As pointed out \cite{soleng1995dark}, this solution reduces to the following particular cases: Schwarzschid-de Sitter solution for $\alpha = -1$; Schwarzschild solution for $\alpha$ = 0 and Reissner-Nordstr\"om solution for $\alpha = 1$. The limit case $\alpha \rightarrow \infty$ results in the Letelier black hole solution with a cloud of strings \cite{letelier1979clouds}.

Using the metric given by Eq. (\ref{eq5}) and considering Eq. (\ref{eq7}), we can calculate the components of the Einstein tensor and, substuting them into the Einsteins fields equations,

\begin{equation} \label{eq8}
G^{\mu}_{\phantom{a}\nu} = 8 \pi T^{\mu}_{\phantom{a}\nu},
\end{equation}

\noindent we obtain

\begin{subeqnarray} \label{eq9}
T^{t}_{\phantom{a}t} &=& T^{r}_{\phantom{a}r} = \frac{\epsilon}{8 \pi r^2} \left( \frac{l}{r}\right)^{2/\alpha},\\
T^{\theta}_{\phantom{a}\theta} &=& T^{\phi}_{\phantom{a}\phi}= - \frac{\epsilon}{8 \pi \alpha r^2} \left( \frac{l}{r}\right)^{2/\alpha}.
\end{subeqnarray}

The solution given by Eq. (\ref{eq7}), with a large value of $\alpha$, represents a fluid of strings radially pointing at low but nonzero temperature \cite{soleng1995dark}. In this scenario, the solution given by Eq. (\ref{eq8}) introduces a $1/r$ correction to Newton's gravitational law which can be used to explain the rotation curves of galaxies.

It is worth calling attention to the fact that Soleng \cite{soleng1995dark} considered the to propose this fluid of strings, in the limit when $\alpha \gg 1$, as a possible source of dark matter. He also considered a cosmological scenario in which case $\alpha \simeq \left(M H_0\right)^{-1/2}$ where $M$ is the mass of the galaxy and $H_0$ is the Hubble constant at present time, and also concluded that it is not possible to use this fluid of strings to solve the missing mass problem, in special, due to the assumption that the gravitational potential energy of the string should be spread over a distance $H_0^{-1}$ in order to get $\alpha \simeq \left(M H_0\right)^{-1}$. It also should be noted that, in a scenario where a phantom field is associated with the dark matter, a similar solution was obtained\cite{li2012galactic} which means that the effects of the fluid of strings considered by Soleng \cite{soleng1995dark} can correspond to similar ones induced bu the phantom fields.

\section{The Reissner-Nordstr\"om-(de Sitter) Anti-de Sitter black hole with a fluid of strings}
\label{sec3}

Now, let us consider a static and spherically symmetric black hole with electric charge $Q$ immersed in a fluid of strings (Reissner-Nordstr\"om black hole with a fluid of strings). The metric corresponding to this system must be a solution of the Einstein equations (Eq. (\ref{eq8})) together with the Maxwell ones which, in the exterior region of the black hole, are given by

\begin{subeqnarray} \label{eq10}
\nabla_\nu F^{\mu \nu} =0, \\
\partial_{[\sigma} F_{\mu \nu]} =0,
\end{subeqnarray}

\noindent whose non-null Maxwell tensor components are given, in the case under consideration, by $F_{10}=-F_{01}=\frac{Q}{\sqrt{4 \pi}r^2}$. The stress-energy tensor associated with this electric charge is given by

\begin{subeqnarray} \label{eq11}
T^{t}_{\phantom{a}t} &=& T^{r}_{\phantom{a}r} = - \frac{Q^2}{8 \pi r^4},\\
T^{\theta}_{\phantom{a}\theta} &=& T^{\phi}_{\phantom{a}\phi}= \frac{Q^2}{8 \pi r^4}.
\end{subeqnarray}

Thus, we can consider that the stress-energy tensor corresponding to the Reissner-Nordstr\"om black hole with a fluid of strings is given by

\begin{subeqnarray} \label{eq12}
T^{t}_{\phantom{a}t} &=& T^{r}_{\phantom{a}r} = - \frac{Q^2}{8 \pi r^4}+\frac{\epsilon}{8 \pi r^2} \left( \frac{l}{r}\right)^{2/\alpha},\\
T^{\theta}_{\phantom{a}\theta} &=& T^{\phi}_{\phantom{a}\phi}= \frac{Q^2}{8 \pi r^4}- \frac{\epsilon}{8 \pi \alpha r^2} \left( \frac{l}{r}\right)^{2/\alpha},
\end{subeqnarray}

\noindent where we have added appropriately the components of the energy-momentum tensor given by Eq. (\ref{eq9}), which corresponds to the case where the electromagnetic field is absent.

Using Einstein's equations and taking into account the metric given by Eq. (\ref{eq5}), we get the following equation

\begin{equation} \label{eq13}
G^{t}_{\phantom{a}t} =\frac{1}{r^2}\left[ (r e^{2 \mu})' -1 \right] = - \frac{Q^2}{r^4}+\frac{\epsilon}{ r^2} \left( \frac{l}{r}\right)^{2/\alpha},
\end{equation}

\noindent where the comma represents the derivative with respect to the coordinate $r$. Integrating the above equation, we obtain the metric corresponding to the Reissner-Nordstr\"om black hole with a fluid of strings, which is given by

\begin{equation} \label{eq14}
ds^2 = - f(r) dt^2 +\frac{1}{g(r)} dr^2+ r^2 d \Omega^2,
\end{equation}

\noindent where

\begin{eqnarray} \label{eq15}
f(r) &=& g(r)= 1- \frac{2M}{r}+\frac{Q^2}{r^2}+ \nonumber \\
&& \left\{
\begin{array}{cc}
\epsilon l r^{-1} \log(\eta r) & \mbox{ for} \quad \alpha =2 \\
\epsilon \alpha( \alpha-2)^{-1} \left( \frac{l}{r}\right)^{2/\alpha} & \mbox{ for} \quad \alpha \neq 2,
\end{array}
\right.
\end{eqnarray}

Finally, if we add the cosmological constant to the obtained metric given by Eq. (\ref{eq13}), we obtain the following result

\begin{eqnarray} \label{eq16}
f(r)&=& 1- \frac{2M}{r}+\frac{Q^2}{r^2}- \frac{\Lambda r^2}{3}+ \nonumber \\
&&\left\{
\begin{array}{cc}
\epsilon l r^{-1} \log(\eta r) & \mbox{ for} \quad \alpha =2 \\
\epsilon \alpha( \alpha-2)^{-1} \left( \frac{l}{r}\right)^{2/\alpha} & \mbox{ for} \quad \alpha \neq 2,
\end{array}
\right.
\end{eqnarray}

The particular cases arising from this metric are the following: $\alpha = -1$ corresponds to the Reissner-Nordstr\"om spacetime with an effective cosmological constant which depends on the parameter $l$; $\alpha = 0$ gives us the Reissner-Nordstr\"om-(de Sitter) anti-de Sitter solution; $\alpha = 1$ represents the RN-anti-de Sitter solution with an effective charge which depends on the parameter $l$; finally, the case $\alpha \rightarrow \infty$, gives us a solution which corresponds to de RN-(de Sitter) anti-de Sitter black hole surrounded by a cloud of strings which is a particular case of a solution obtained recently \cite{toledo2019reissner} by removing the contribution of the quintessence field.

For the case under consideration, the components of the Einstein tensor can be written as

\begin{subeqnarray} \label{eq17}
G_{tt} &=& - f(r) \left[ - \frac{Q^2}{ r^4}+\frac{\epsilon}{ r^2} \left( \frac{l}{r}\right)^{2/\alpha}\right] - \Lambda f(r), \\
G_{rr}& =& \frac{1}{f(r)} \left[ - \frac{Q^2}{ r^4}+\frac{\epsilon}{ r^2} \left( \frac{l}{r}\right)^{2/\alpha}\right] +\frac{ \Lambda}{ f(r)},\\
G_{\theta \theta} &=& G^{\phi}_{\phantom{a}\phi}= r^2\left[ \frac{Q^2}{r^4}- \frac{\epsilon}{\alpha r^2} \left( \frac{l}{r}\right)^{2/\alpha}+ \Lambda \right],\\
G_{\phi \phi}&=& r^2 \sin^2 \theta \left[ \frac{Q^2}{r^4}- \frac{\epsilon}{\alpha r^2} \left( \frac{l}{r}\right)^{2/\alpha}+ \Lambda \right].
\end{subeqnarray}

As a consequence, we can conclude that Eq. (\ref{eq13}), in which $f(r)$ is given by Eq. (\ref{eq16}), is the metric corresponding to the Reissner-Nordstr\"om-AdS black hole with a fluid of strings, which is a solution of the Einsteins equations with cosmological constant, namely

\begin{equation} \label{eq18}
G_{\mu \nu} = R_{\mu \nu} - \frac{1}{2} Rg_{\mu \nu} + \Lambda g_{\mu \nu} = 8 \pi T_{\mu \nu}.
\end{equation}

\subsection{Black hole horizons}

The black hole horizons, in the considered coordinate system, can be defined by the equation

\begin{equation}
f(r) = 0.
\end{equation}

\begin{figure*}[!htb]
\centering
\includegraphics[scale=0.5]{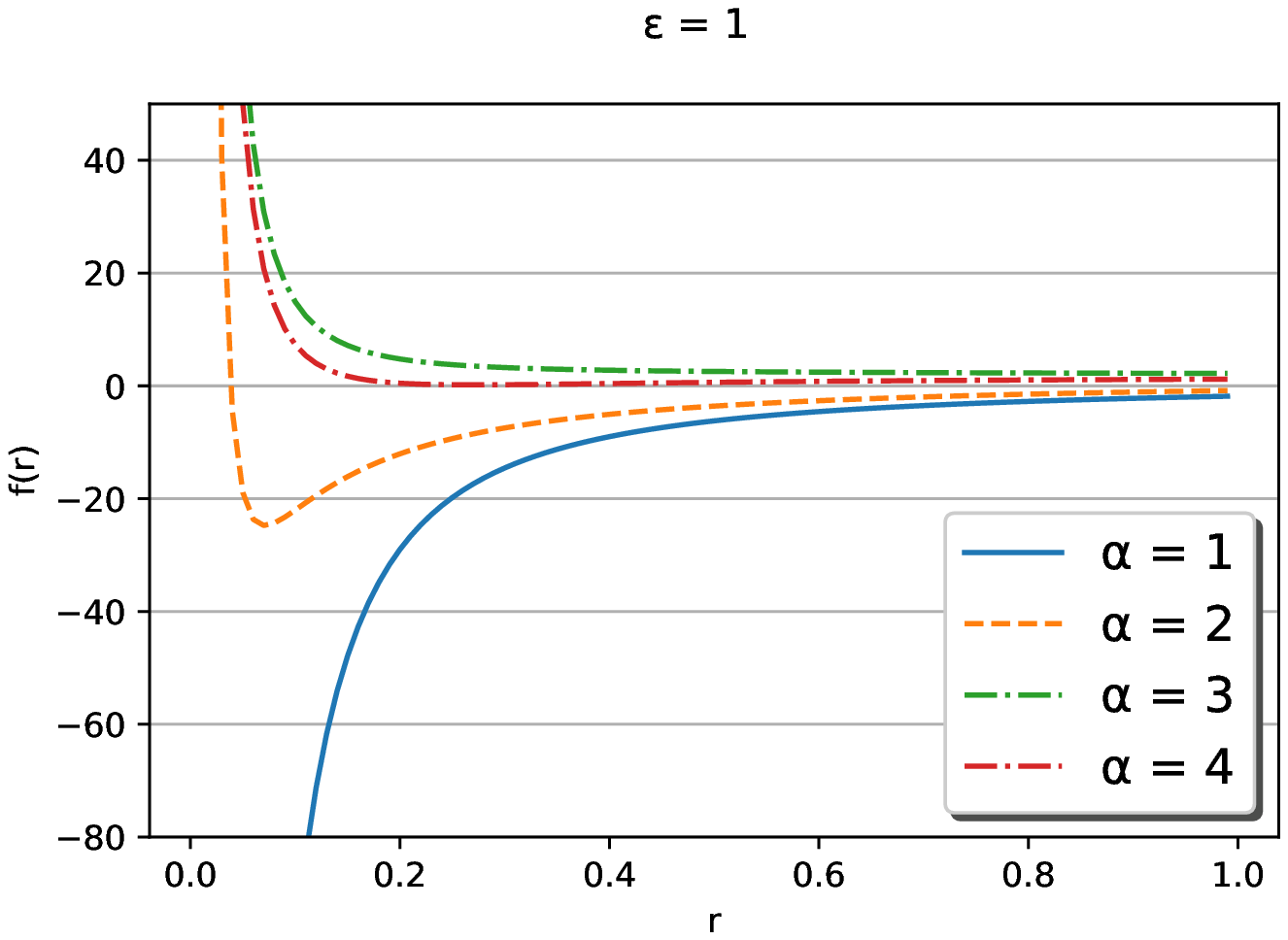}
\includegraphics[scale=0.5]{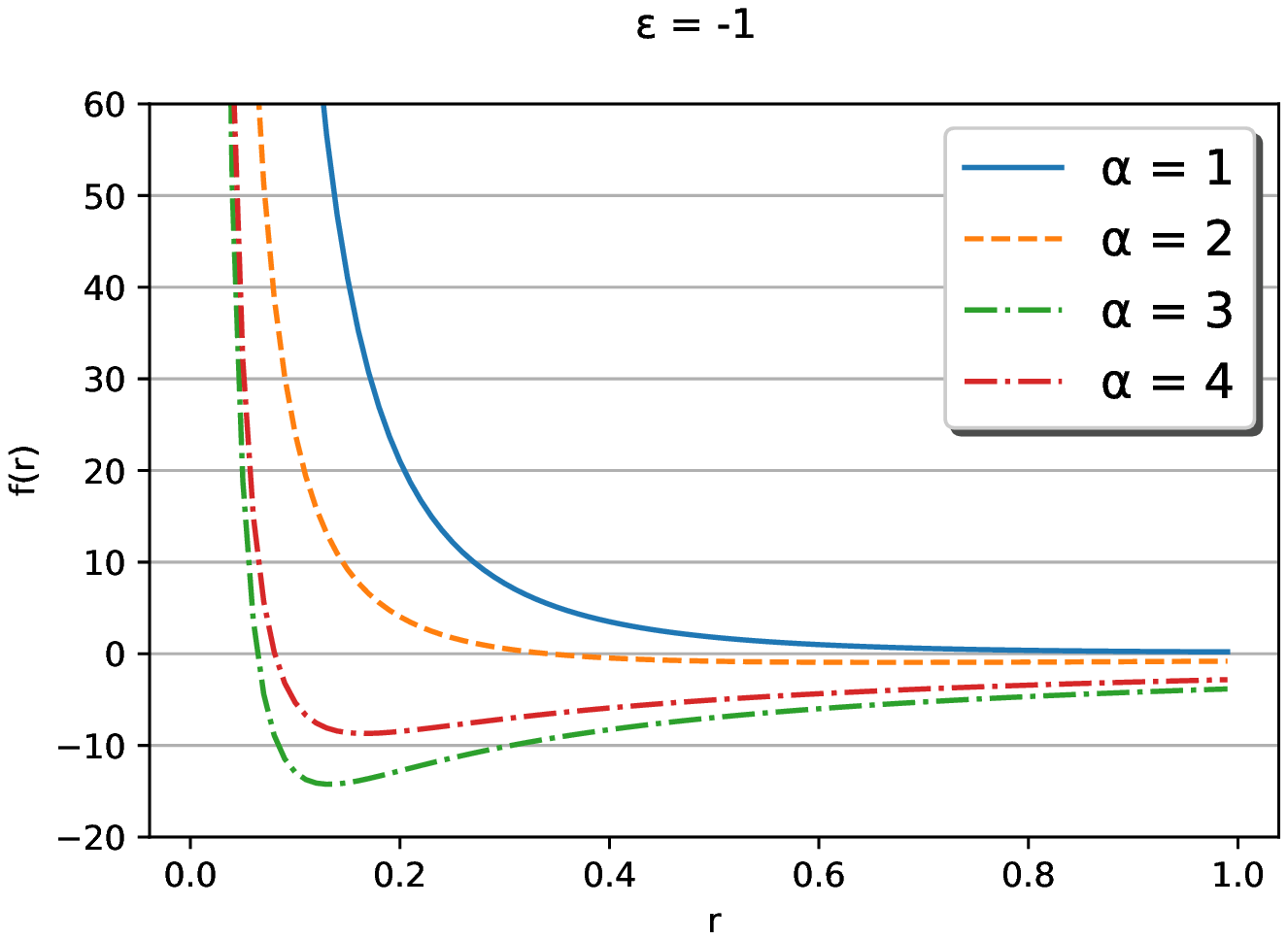}
\caption{$f(r)$ for $ M = l =\eta = 1$, $Q^2 = 0.2$ and different values of $\alpha$.}
\label{Fig1}
\end{figure*}

\begin{figure*}[!htb]
\centering
\includegraphics[scale=0.5]{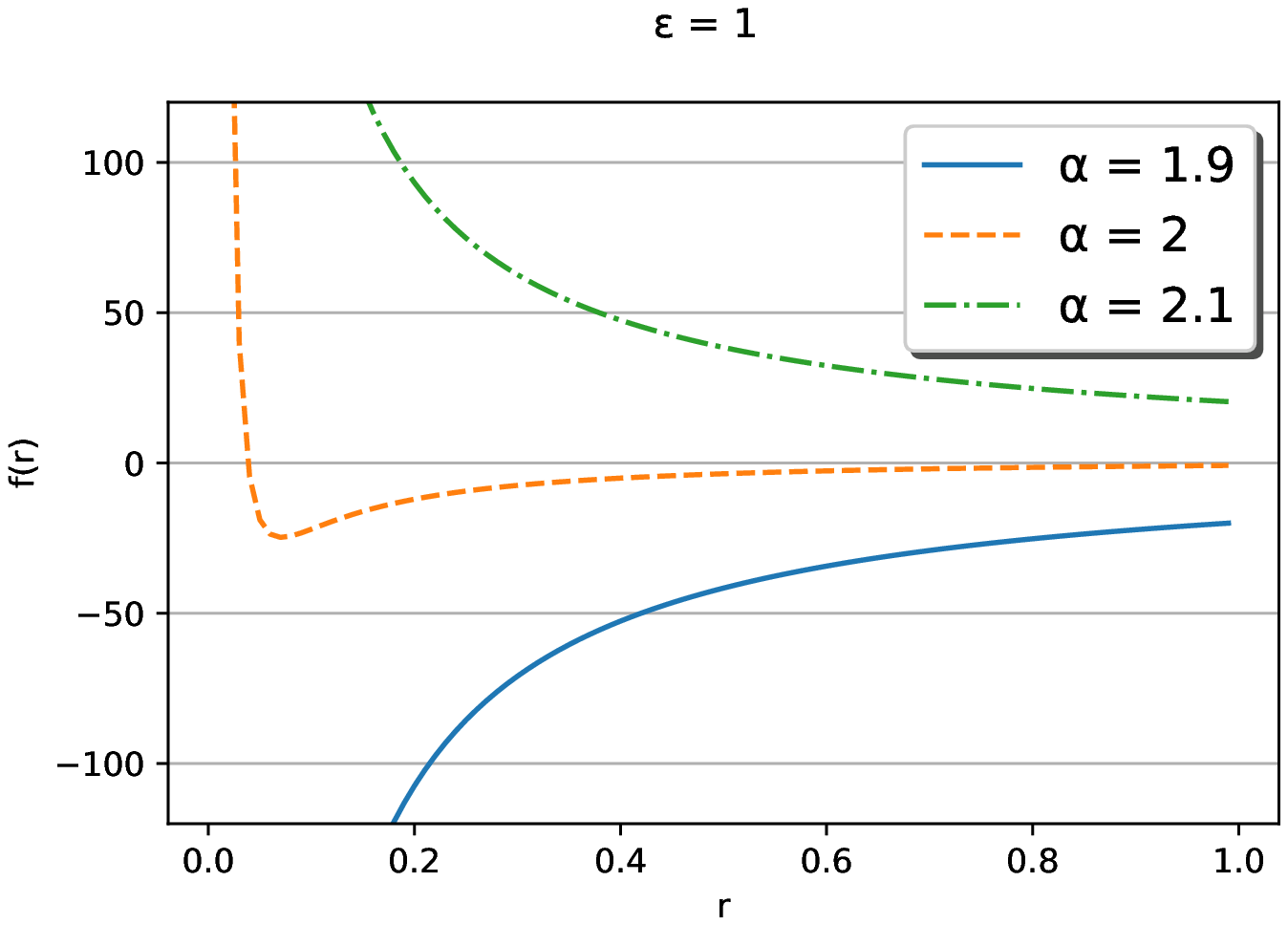}
\includegraphics[scale=0.5]{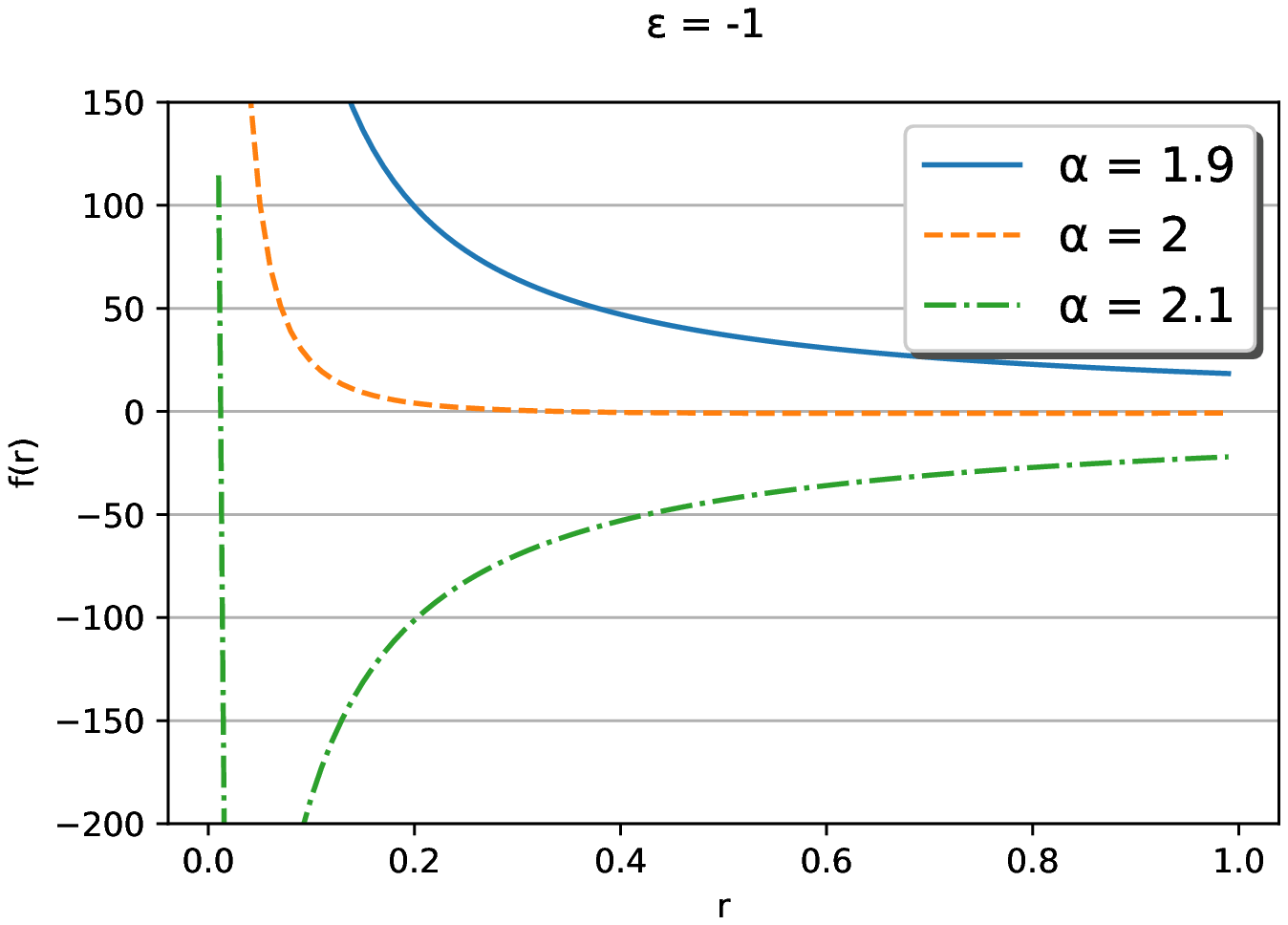}
\caption{$f(r)$ for $ M = l = 1$, $Q^2 = 0.2$ and for values of $\alpha$ near $\alpha = 2 $.}
\label{Fig2}
\end{figure*}

In Figs. \ref{Fig1} and \ref{Fig2}, we represent the function $f(r)$ for different values of the parameters associated with the fluid of strings. The roots of the function $f(r)$ determine the black hole horizons.

In Fig. \ref{Fig1} and \ref{Fig2}, we show the behaviors of $f(r)$ for different values of $\alpha$. In the left panel, in both figures, the signal of $\epsilon$ is positive, while in the right panel it is negative.

Due to the small value of the cosmological constant, we can verify that the black hole has always a horizon very far from the singularity, which is not represented in the figures. Thus, we can conclude that the system has until three positive horizons: a cosmological horizon associated with the cosmological constant ($r_c$), the event horizon ($r_+$) and the interior horizon ($r_-$).

\subsection{Geodesic motion}
\label{sec4}

Let us consider the movement of a test particle with mass $m$ and electric charge $q$ around a black hole described by the metric given in Eq. (\ref{eq13}) with $f(r)$ given by Eq. (\ref{eq16}). We will neglect the effect of the cosmological constant, because its value is very small when if compared with the other parameters, only having importance on the cosmological scale.

Given a Killing field, $\xi^\mu$, the quantity $\xi^\mu(p_\mu+q A_\mu)$, where $p^\mu = m u^\mu = m dx^\mu/d \tau $ is the 4-momentum of the particle and $A_\mu$ is the electromagnetic 4-potencial, is conserved. Taking into account the timelike Killing vector $\xi^\mu = (1,0,0,0)$, we conclude that the energy

\begin{equation}
E = - p_0 - q A_0 = - p_0 +\frac{qQ}{r}
\end{equation}

\noindent is conserved. Due to the symmetry of the problem and the azimuthal component of the Killing field, we conclude that the angular momentum is given by

\begin{equation}
L = p_\phi.
\end{equation}

We can conclude, also, that the movement of the test particle around the black hole occurs in a plane. Thus, we get $p_\theta = 0$. Considering that the radial velocity of the particle is null, the equation $p_\mu p ^\mu = - m^2$ implies

\begin{equation}
\left(E- \frac{qQ}{r} \right)^2 =m^2 \left( \frac{dr}{d \tau}\right)^2 + \left( m^2 - \frac{L^2}{r^2} \right) f(r).
\end{equation}

This equation can be written in the form

\begin{equation}
E = m+\frac{m}{2} \left( \frac{dr}{d \tau}\right)^2+U_{eff},
\end{equation}

\noindent where

\begin{eqnarray} \label{eq15}
U_{eff} &\approx& \frac{qQ}{r} - \frac{mM}{r} + \frac{L^2}{2 m r^2} \nonumber \\
&&+\left\{
\begin{array}{cc}
\frac{m \epsilon l}{2r} \log(\eta r), & \mbox{ for} \quad \alpha =2 \\
\frac{m \epsilon \alpha}{2( \alpha-2)} \left( \frac{l}{r}\right)^{2/\alpha}, & \mbox{ for} \quad \alpha \neq 2.
\end{array}
\right.
\end{eqnarray}

\begin{figure*}[!htb]
\centering
\includegraphics[scale=0.7]{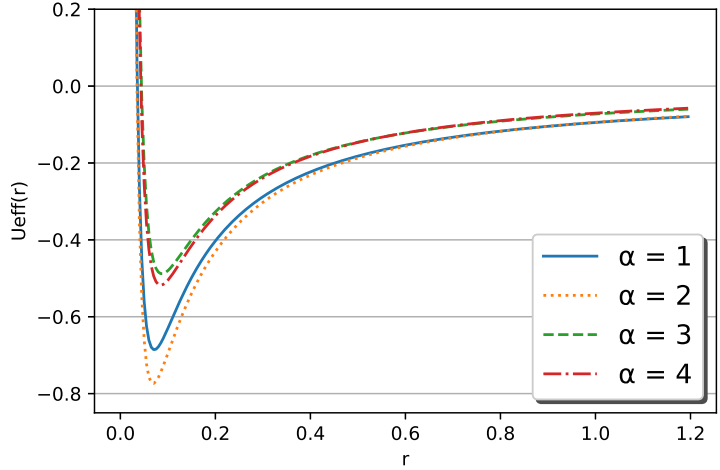}
\caption{The effective potential $U_{eff}(r)$ for different values of $\alpha$.}
\label{Fig3}
\end{figure*}

We considered approximations up to the order $1/r^2$. Depending on the values of the parameters, $U_{eff}$ has a minimum, as can be seen in Fig (\ref{Fig3}). At this point, the movement of the particle describes a circle around the black hole.

At the minimum of $U_{eff}$, we get

\begin{eqnarray} \label{eq15}
\frac{d U_{eff}}{dr} &=& 0 = -\frac{qQ}{r^2} + \frac{mM}{r^2} - \frac{L^2}{m r^3} \nonumber \\
&&-\left\{
\begin{array}{cc}
\frac{m \epsilon l}{2r^2} \left[ \log(\eta r)-1\right]& \mbox{ for} \quad \alpha =2 \\
\frac{m \epsilon}{( 2- \alpha)} \left( \frac{l}{r}\right)^{2/\alpha-1} & \mbox{ for} \quad \alpha \neq 2,
\end{array}
\right.
\end{eqnarray}

Let us consider $v$ the modulus of the velocity of the particle in its orbit, thus we can write $L^2 = m^2 v^2 r^2$, and therefore

\begin{equation} \label{eq15}
v^2 \sim - \frac{ qQ}{ m r} + \frac{M}{ r} -\left\{
\begin{array}{cc}
\frac{ \epsilon l}{2r} \left[ \log(\eta r)-1\right]& \mbox{ for} \quad \alpha =2 \\
\frac{r \epsilon}{( 2- \alpha)} \left( \frac{l}{r}\right)^{2/\alpha-1} & \mbox{ for} \quad \alpha \neq 2,
\end{array}
\right.
\end{equation}

\begin{figure*}[!htb]
\centering
\includegraphics[scale=0.7]{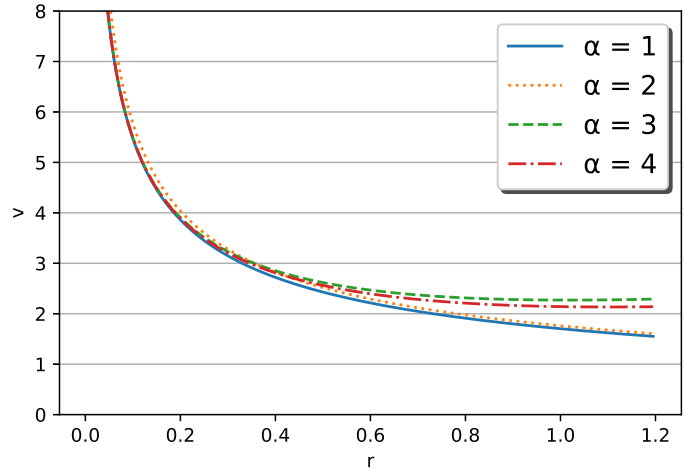}
\caption{The velocity $v$ for different values of the parameter $\alpha$.}
\label{Fig4}
\end{figure*}

The behavior of $v$ as a function of $r$ is represented in Fig. \ref{Fig4} for different values of the parameter $\alpha$, covering the two different solutions, for $\alpha = 2$ and $\alpha \neq 2$.

From Fig. \ref{Fig4} we can note that, depending on the values of the parameter $\alpha$, the values of $v$ do not tend to zero, as should be expected in the absence of the fluid of strings. Note that in all cases the velocity decreases, and decreases more with the decreasing of the parameter.

\section{Kerr-Newman black hole with a fluid of strings}
\label{sec4}

In this section, we obtain the metric corresponding to a rotating charged black hole immersed in a fluid of strings (Kerr-Newman black hole surrounded by a fluid of strings). To obtain the correct metric, we use the Newman-Janis method \cite{newman1965note, newman1965metric}, with the adaptation proposed by Azreg-A\"inou \cite{azreg2014generating}. This modified method can generate regular stationary solutions for black holes, considering the physical properties and the problem of symmetry \cite{azreg2014generating}.

Firstly, let us consider the spacetime line element of a static charged black hole with a fluid of strings surrounding it, which can be written as \cite{letelier1979clouds}

\begin{equation} \label{nj14}
ds^2= f(r) dt^2 - \frac{1}{g(r)}dr^2 - h^2 d \Omega^2,
\end{equation}

\noindent where $f(r) = g(r)$ as explicitly given by Eq. (\ref{eq15}) taking $\Lambda =0$.

In order to write the metric given by Eq. (\ref{nj14}) into the Eddington-Finkelstein coordinates, let us use the coordinate transformation \cite{azreg2014generating}

\begin{equation} \label{nj15}
du = dt-\frac{dr}{\sqrt{fg}},
\end{equation}

\noindent and, thus, we get

\begin{equation} \label{nj16}
ds^2 = f du^2- \sqrt{\frac{f}{g}} du dr - h^2 d\Omega^2.
\end{equation}

Now, we determine the null tetrad basis that describes the metric, by the equation

\begin{equation} \label{nj3}
g^{\mu \nu} = l^{\mu} n^{\nu} + l^{\nu} n^{\mu} - m^{\mu}\bar{m}^{\nu} - m^{\nu} \bar{m}^{\mu}
\end{equation}

\noindent where $l$ and $n$ are real numbers and $\bar{m}$ is the complex conjugate of $m$. Beside that, the tetrades obey the relations $l_{\mu}l^{\nu}=m_{\mu} m^{\mu}= n_{\nu} n^{\nu} =0$, $l_{\mu} m^{\mu} = n_{\mu} m^{\mu} = 0$ e $l_{\mu} n^{\mu} = - m_{\mu} \bar{m}^{\mu} = 0$.

Using the metric given by Eq. (\ref{nj16}), we get

\begin{eqnarray} \label{nj17}
l^\mu &=& \delta^{\mu}_{1}, \nonumber \\
n^\mu &=& \sqrt{\frac{g}{f}} \delta^{\mu}_{0} - \frac{f}{2} \delta^{\mu}_{1},\nonumber \\
m^\mu &=& \frac{1}{\sqrt{2} h} \left( \delta^{\mu}_{2} + \frac{i}{\sin \theta} \delta^{\mu}_{3} \right), \nonumber \\
\bar{ m}^\mu &=& \frac{1}{\sqrt{2} h} \left( \delta^{\mu}_{2} - \frac{i}{\sin \theta} \delta^{\mu}_{3} \right).
\end{eqnarray}

Now, we perform the transformation $u \rightarrow u - ia \cos \theta$ and $r \rightarrow r - ia \cos \theta$ \cite{azreg2014generating}, where $a$ is a parameter associated to the rotation of the source. We also do the changes $f(r) \rightarrow F(r,a, \theta)$, $g(r) \rightarrow G(r,a, \theta)$ and $h \rightarrow \Sigma(r,a, \theta)$. As a consequence, the null tetrades are written in the form \cite{azreg2014generating}

\begin{eqnarray} \label{nj18}
l^\mu &=& \delta^{\mu}_{1}, \nonumber \\
m^\mu &=& \frac{1}{\sqrt{2} \Sigma} \left[ \delta^{\mu}_{2} +ia \sin \theta (\delta^\mu_0-\delta^\mu_1)+ \frac{i}{\sin \theta} \delta^{\mu}_{3} \right]\nonumber \\
n^\mu &=& \sqrt{\frac{G}{f}} \delta^{\mu}_{0} - \frac{F}{2} \delta^{\mu}_{1}, \nonumber \\
\bar{ m}^\mu &=& \frac{1}{\sqrt{2} \Sigma} \left[ \delta^{\mu}_{2} -ia \sin \theta (\delta^\mu_0-\delta^\mu_1)- \frac{i}{\sin \theta} \delta^{\mu}_{3} \right].
\end{eqnarray}

The components of the metric in the Eddington-Finkelstein coordinates are, then, given by

\begin{eqnarray} \label{nj19}
g_{00} &=& -F, \nonumber \\
g_{01}&=& g_{10} = - \sqrt{\frac{G}{F}}, \nonumber \\
g_{22} &=& \Sigma^2, \nonumber \\
g_{33} &=& \sin^2 \theta \left[ \Sigma^2+ a^2 \left( 2 \sqrt{\frac{F}{G}} -F \right)\sin^2 \theta \right], \nonumber \\
g_{03}&=& g_{30} = a \left( F - \sqrt{\frac{F}{G}} \right)\sin^2 \theta, \nonumber \\
g_{13}&=& g_{31} =a \sin^2 \theta \sqrt{\frac{F}{G}}.
\end{eqnarray}

Finally, we write the metric in the Boyer-Lindquist coordinates using the transformations

\begin{equation} \label{nj20}
du \rightarrow dt + \lambda (r) dr, \quad d \phi \rightarrow d \phi + \chi (r) dr,
\end{equation}

\noindent where

\begin{subeqnarray} \label{nj21}
\lambda (r) = - \frac{k(r)+a^2}{g(r) h(r) +a^2},\\
\chi (r) = - \frac{a}{g(r) h(r) +a^2},\\
k(r) = \sqrt{\frac{g(r)}{f(r)}} h(r)
\end{subeqnarray}

\noindent are chosen such that the non-diagonal components of the metric are null, excepting $g_{03}$ and $g_{30}$. We can also write

\begin{eqnarray} \label{nj22}
F(r, \theta) &=& \frac{(g h a^2 \cos^2 \theta )}{(k +a^2 \cos^2 \theta)^2}, \nonumber \\
G(r,\theta) &=& \frac{gh+ a^2 \cos^2 \theta}{\Sigma},
\end{eqnarray}

\noindent with $\Sigma$ being given by

\begin{equation} \label{sigma}
\Sigma = r^2 + a^2 \cos^2 \theta.
\end{equation}

In the case under consideration, due to the symmetry, we can use $f(r) = g(r)$, $h(r) = r^2$ and $k(r)=h(r)$. Lastly, we obtain the metric

\begin{eqnarray} \label{metrickn}
ds^2 &=& - \frac{\Delta - a^2 \sin^2 \theta}{\Sigma}dt^2 \nonumber +\frac{\Sigma}{\Delta}dr^2 \nonumber \\
&&- 2 a \sin^2 \theta \left(1 - \frac{\Delta - a^2 \sin^2 \theta}{\Sigma} \right) dt d\phi - \Sigma d \theta^2 \nonumber \\
&&+\sin^2 \theta \left[\Sigma + a^2 \sin^2 \theta \left( 2 - \frac{\Delta - a^2 \sin^2 \theta}{\Sigma} \right) \right] d\phi^2,
\end{eqnarray}

\noindent with

\begin{eqnarray} \label{delta}
\Delta&=& r^2 +a^2 +Q^2 - 2 M r \nonumber \\
&&+\left\{
\begin{array}{cc}
\epsilon l r \log(\eta r) & \mbox{ for} \quad \alpha =2 \\
\epsilon \alpha( \alpha-2)^{-1} r^2 \left( \frac{l}{r}\right)^{2/\alpha} & \mbox{ for} \quad \alpha \neq 2,
\end{array}
\right.
\end{eqnarray}

This metric corresponds to the one that describes a charged, rotating black hole surrounded by a fluid of strings in spherical symmetry.

\subsection{Black hole horizons}

The horizons of a rotating black hole are determined by the equation $g^{rr} = 0$. Using the metric given by Eq. (\ref{metrickn}), the black hole horizons are determined by the equation

\begin{eqnarray} \label{hor}
0 &=& r^2 +a^2 +Q^2 - 2 M r+\nonumber \\
&&\left\{
\begin{array}{cc}
\epsilon l r \log(\eta r) & \mbox{ for} \quad \alpha =2 \\
\epsilon \alpha( \alpha-2)^{-1} r^2 \left( \frac{l}{r}\right)^{2/\alpha} & \mbox{ for} \quad \alpha \neq 2,
\end{array}
\right.
\end{eqnarray}

\begin{figure*}[!htb]
\centering
\includegraphics[scale=0.5]{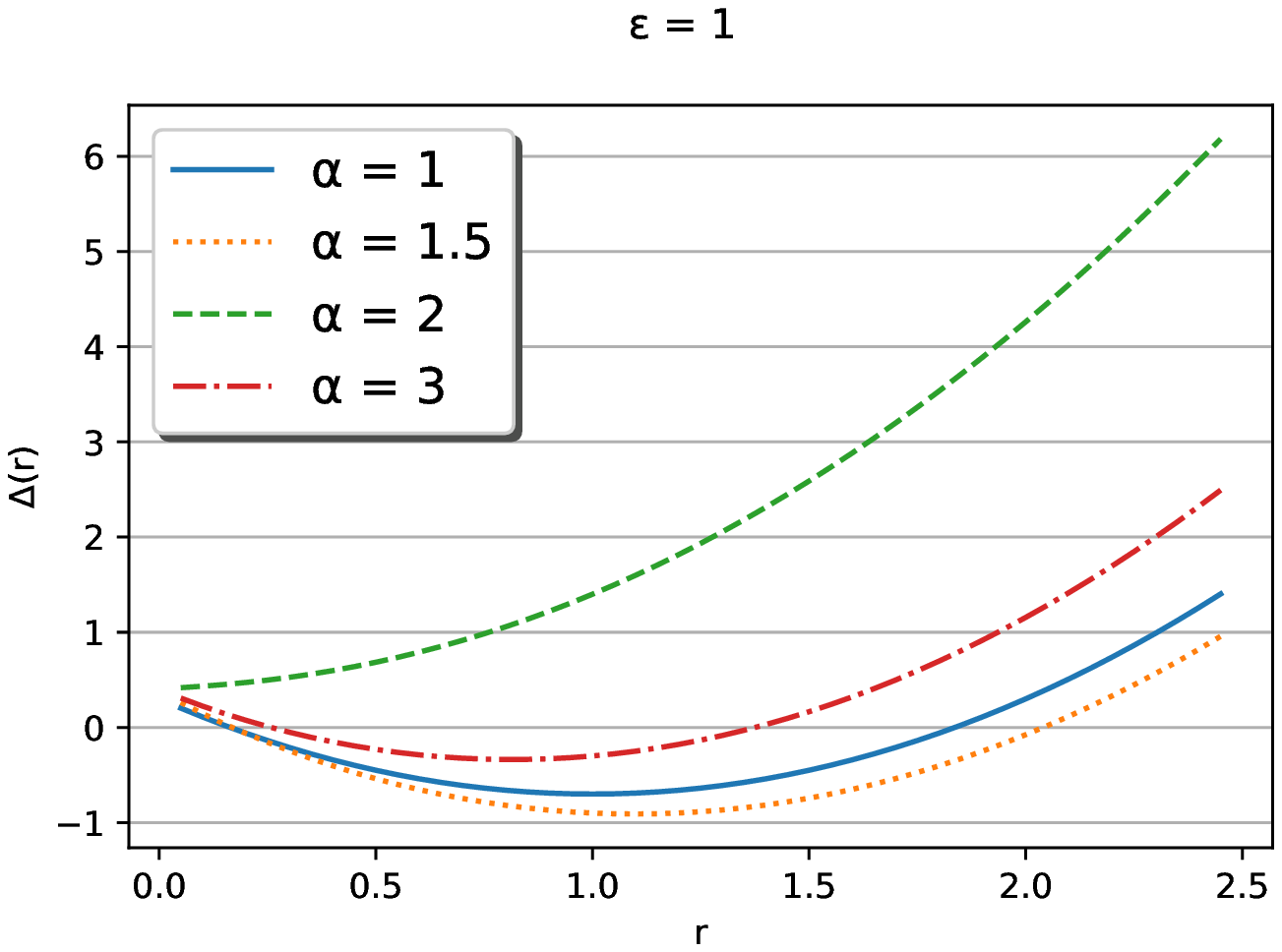}
\includegraphics[scale=0.5]{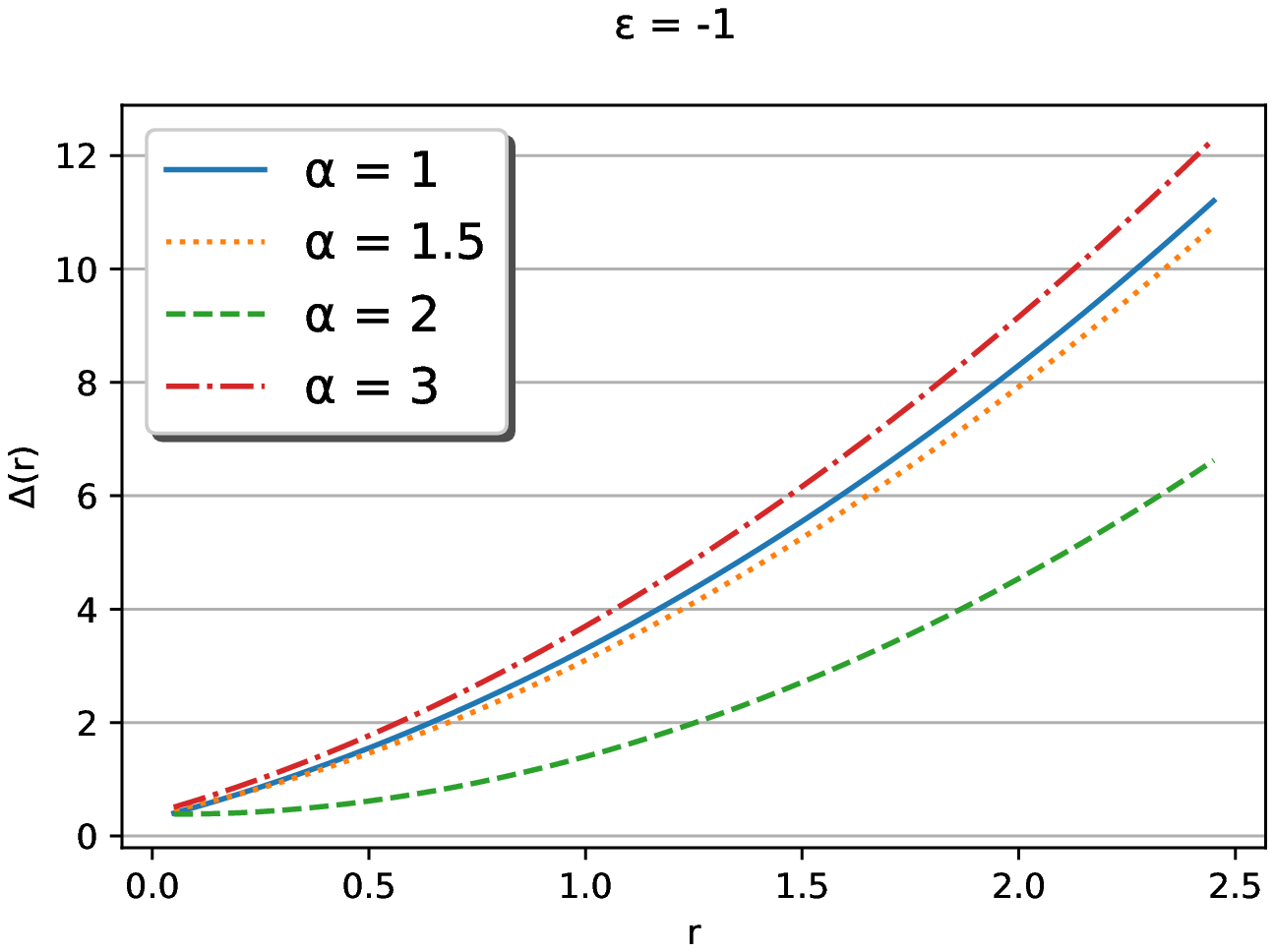}
\caption{The function $\Delta(r)$ for $ M = l =\eta = 1$, $Q^2 = 0.2$ and different values of $\alpha$.}
\label{Fig5}
\end{figure*}

In Fig. \ref{Fig5}, we represent $\Delta(r)$ for different values of the parameter $\alpha$. From Eq. (\ref{hor}), we can observe that the black hole horizons are obtained by the zeros of the function $\Delta(r)$ and, from Fig. \ref{Fig5}, we can notice that the number of horizons of the black hole depends on the values of the parameter associated with the fluid of strings.

Note that, for $\alpha = 2$, the behavior is slightly different as compared with the other values of $a$, as shown in the left panel of Fig. \ref{Fig5}. In the right panel, the behavior is similar for different values of $\alpha$, including $\alpha = 2$.

\subsection{Static surfaces and ergoregions}

The static surfaces of a rotating black hole are determined by $g_{tt}= 0 $. From the metric given by Eq. (\ref{metrickn}), we get

\begin{equation}
\Delta = a^2 \sin^2 \theta.
\end{equation}

\begin{figure*}[!htb]
\centering
\includegraphics[scale=0.5]{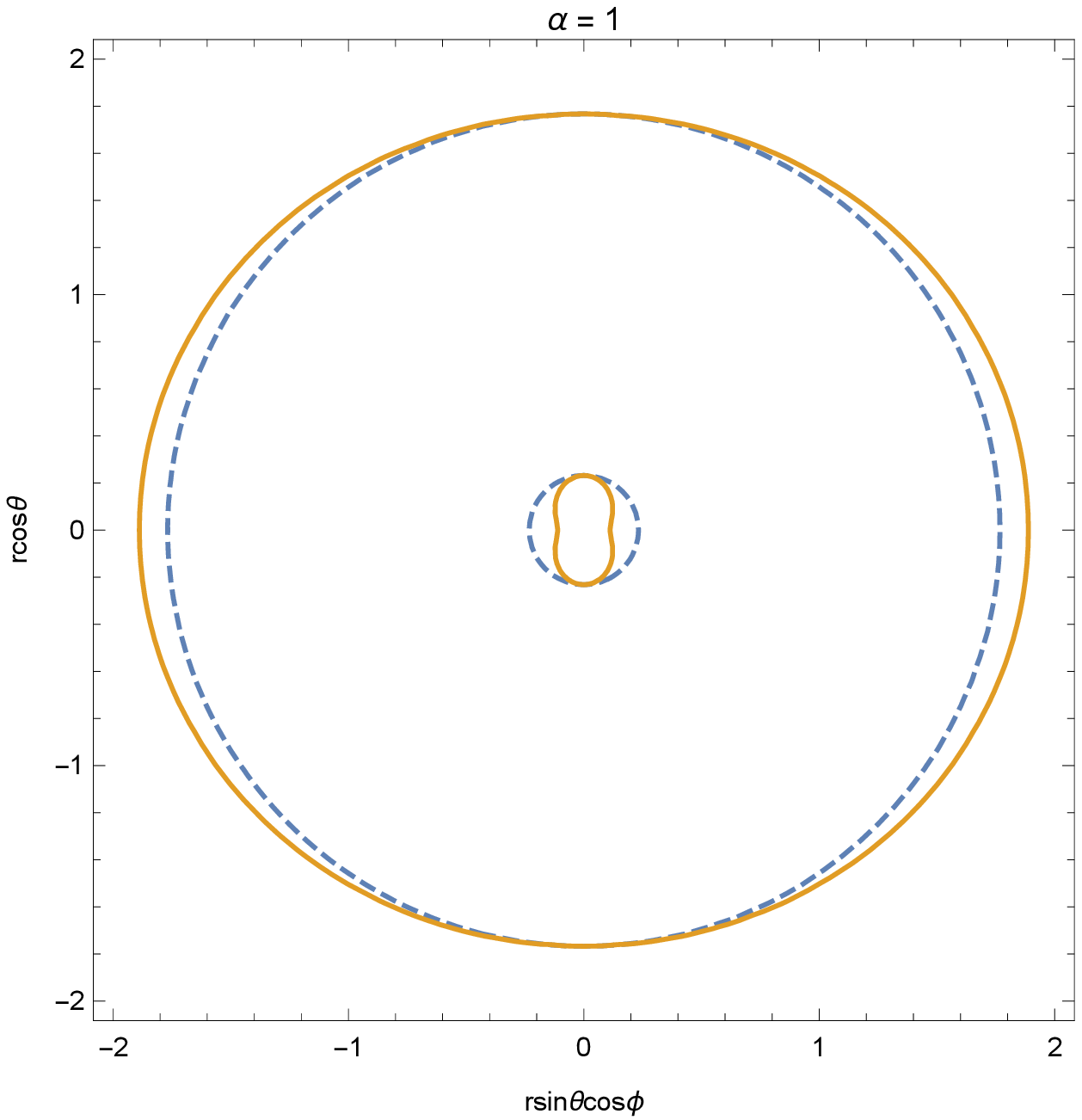}
\includegraphics[scale=0.5]{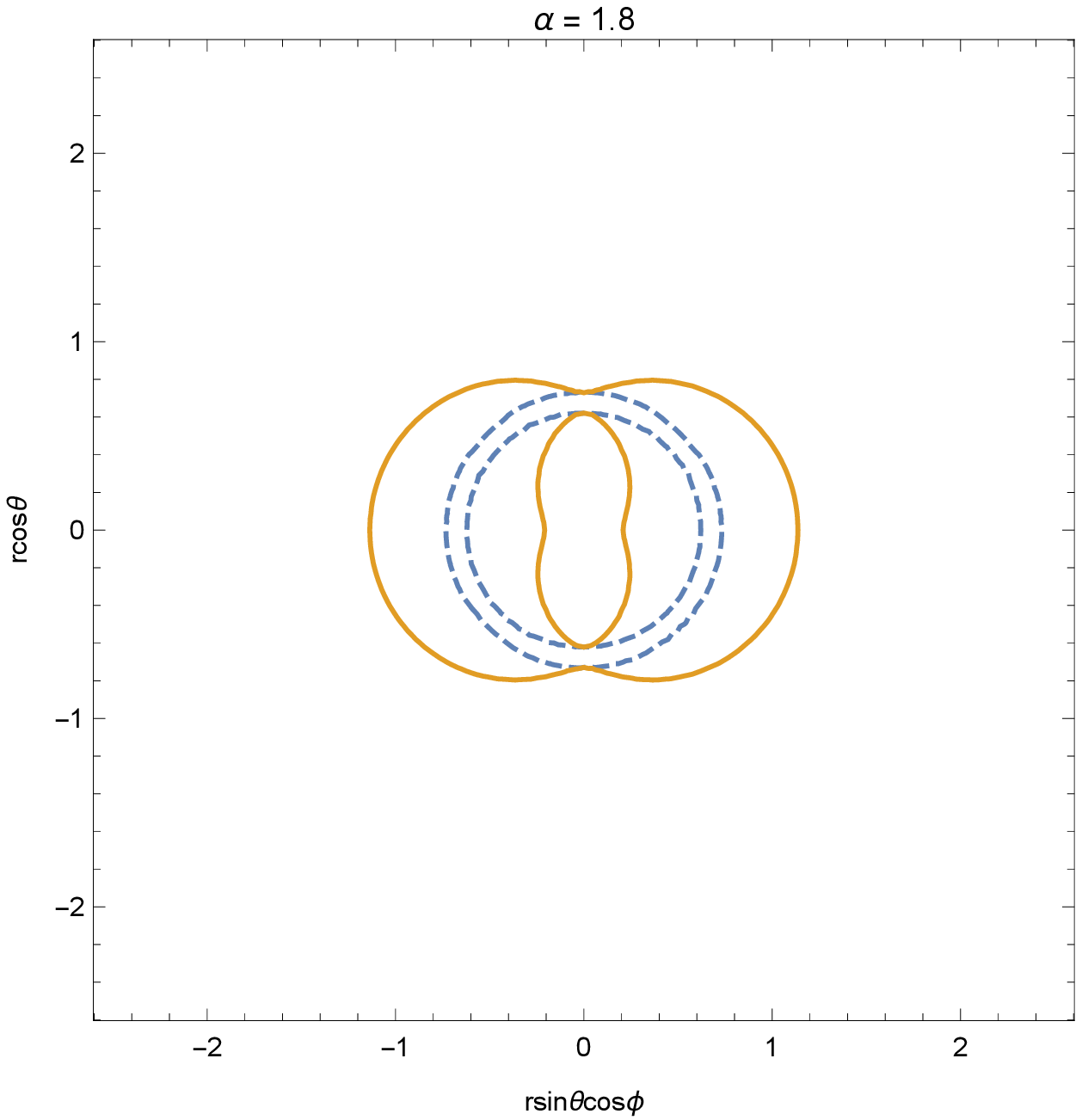}
\includegraphics[scale=0.5]{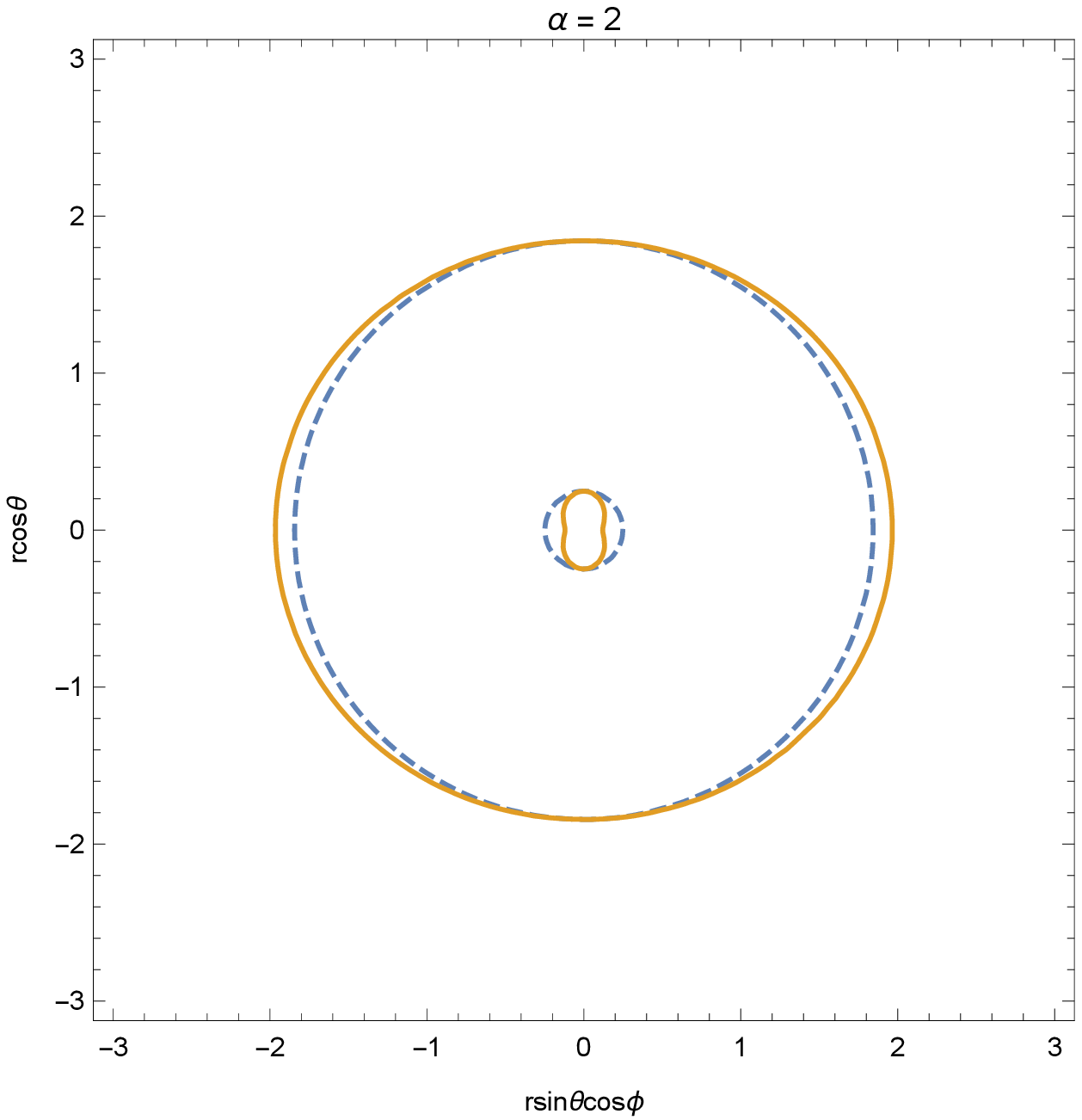}
\includegraphics[scale=0.5]{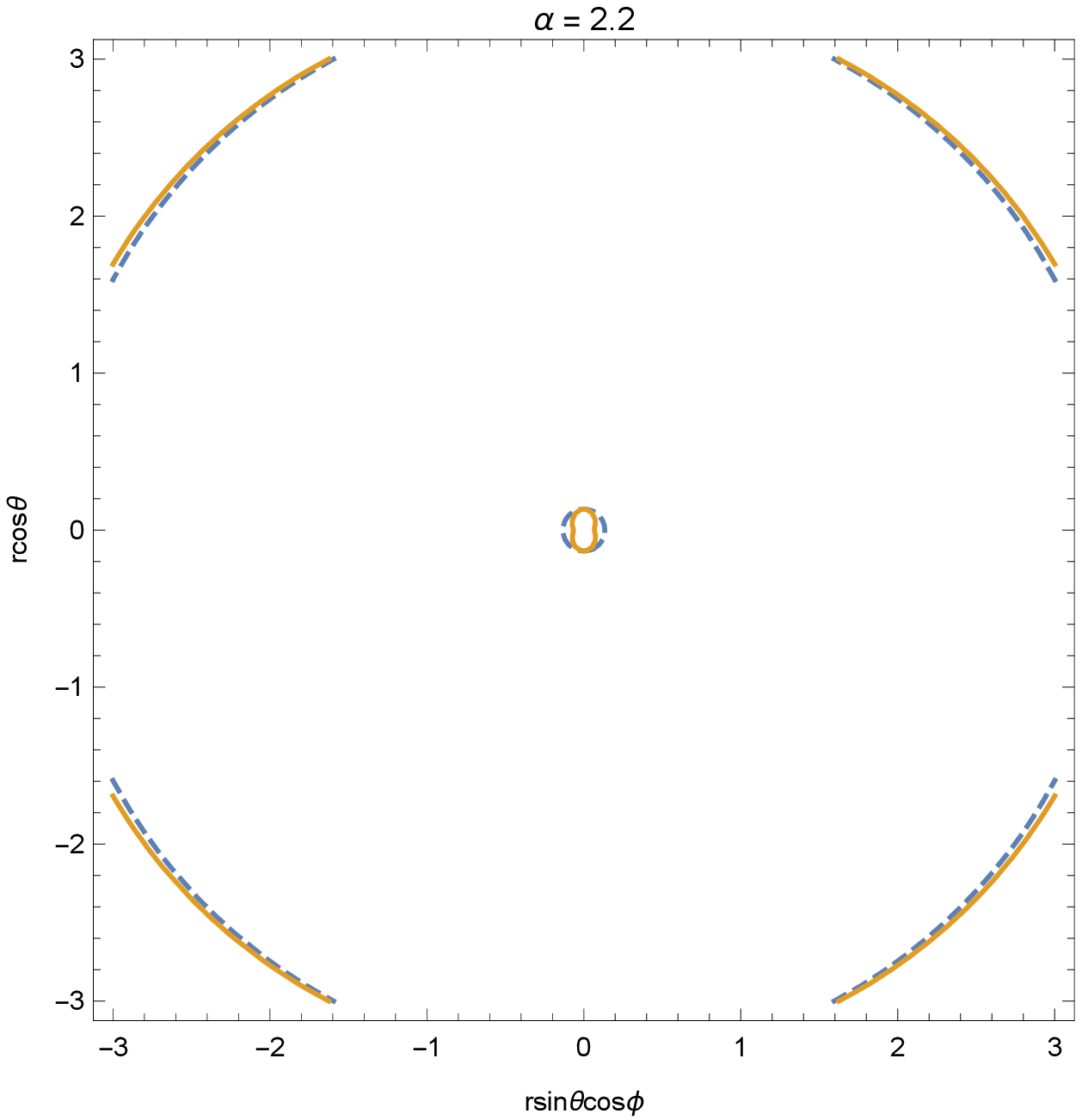}
\includegraphics[scale=0.5]{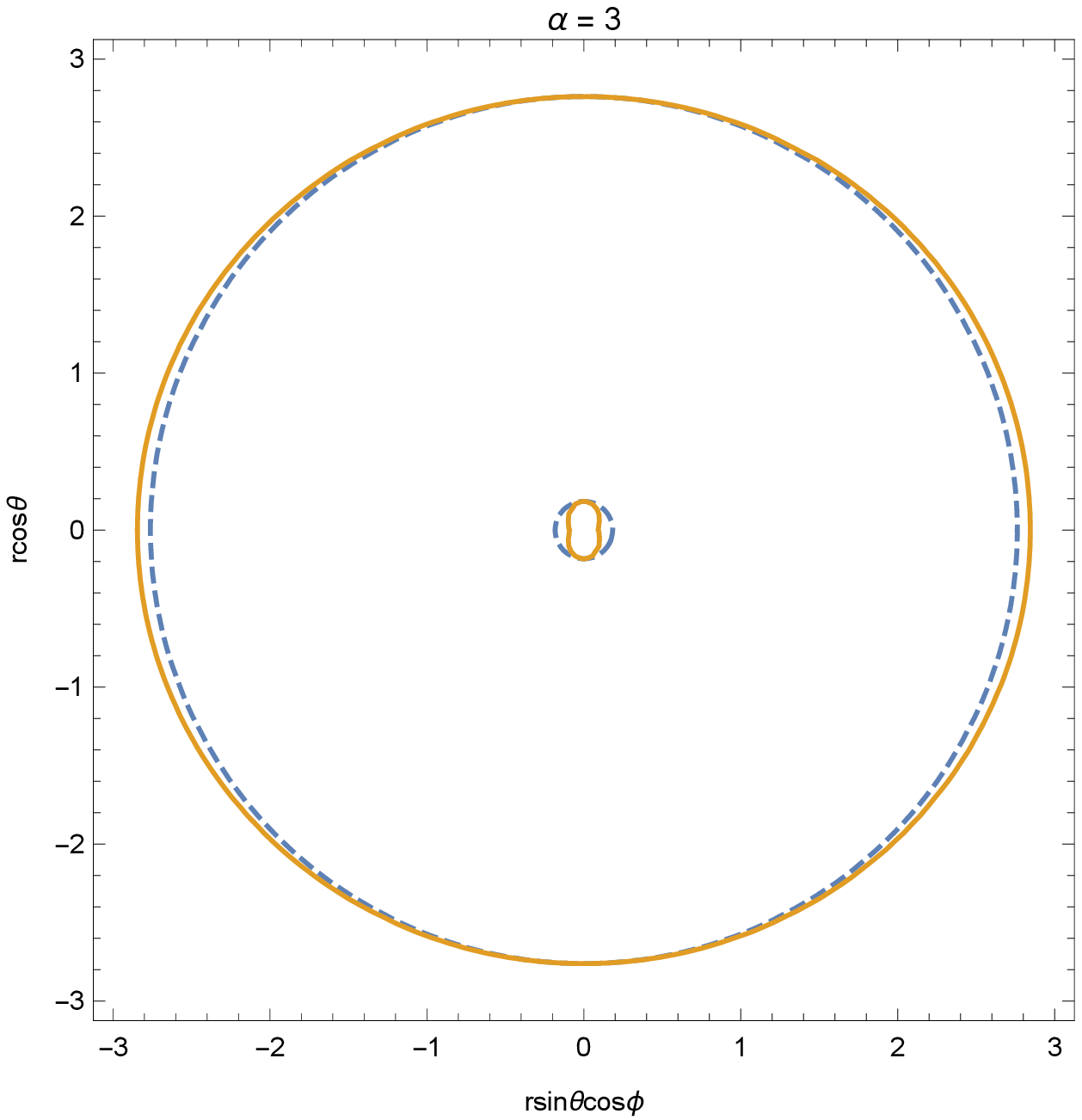}
\includegraphics[scale=0.5]{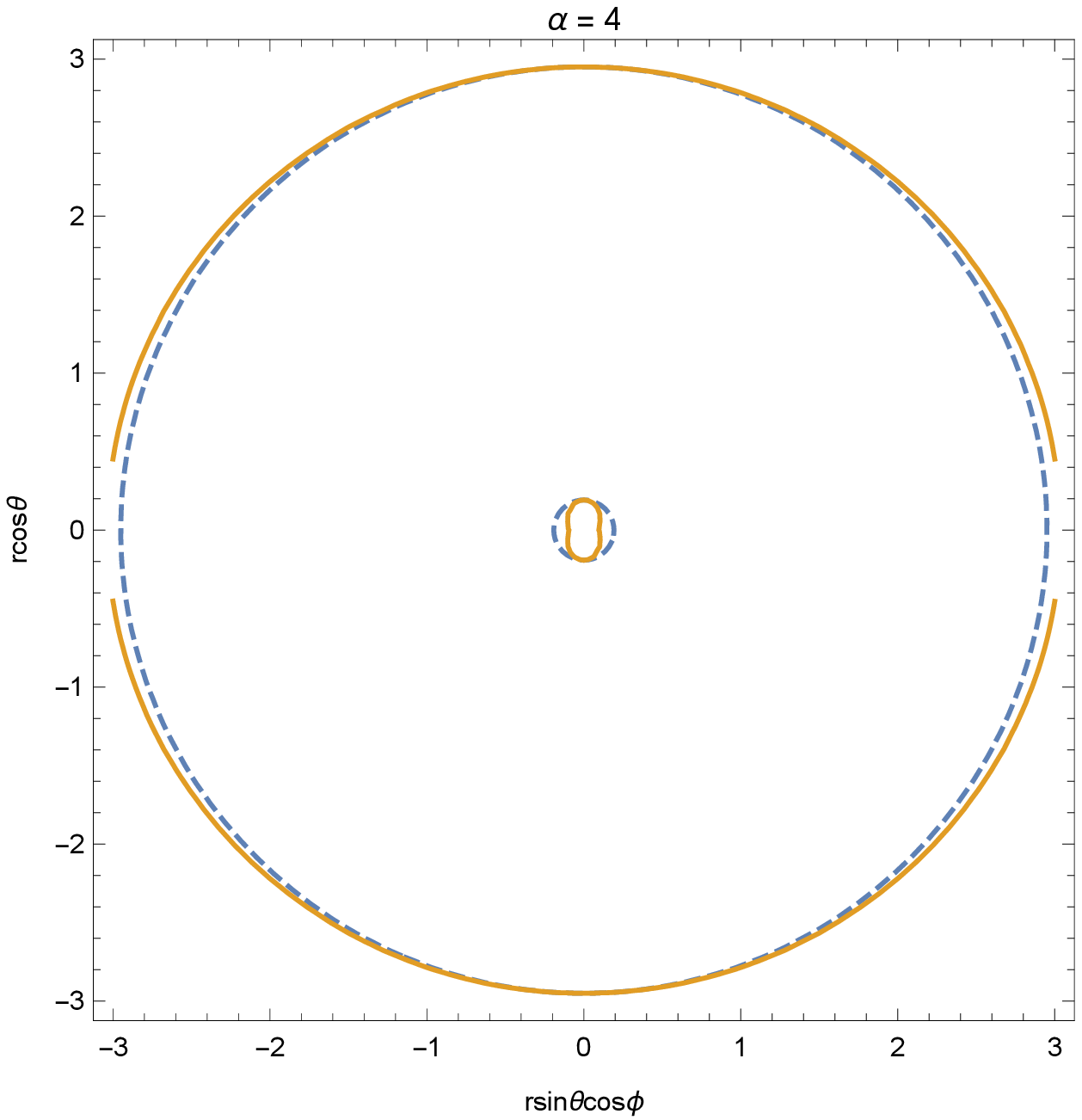}
\caption{Ergoregions. The dashed and the solid lines represent, respecrively, the black hole horizons and the static surfaces.}
\label{Fig6}
\end{figure*}

The ergoregion or ergosphere is the region between the black hole horizon and the static surface and plays an important role in general relativity: an observer in this region cannot remain stationary \cite{misner2017gravitation,wald1984}. In Fig. \ref{Fig6}, we display the ergoregions in the metric under consideration.

\subsection{Geodesic motion}

Nearby a black hole, the movement of a massive particle or a massless one, like a photon, can be described by the Hamiltonian \cite{schee2016silhouette}

\begin{equation} \label{ecom30}
H = g^{\mu \nu} p_\mu p_\nu,
\end{equation}

\noindent where $p^\mu$ are the components of the 4-momentum and $H = 0 $ for photons or $H = - \frac{1}{2} m^2$ for a particle with mass $m$. Due to the symmetries of the system under consideration, we can define two constants of motion: the energy, $E = -p_t$, and the angular momentum, $L=-p_\phi$ related to the axial symmetry. Using the Hamiltonian given by Eq. (\ref{ecom30}) and the Hamilton-Jacobi equations, we get the following set of differential equations \cite{toshmatov2017rotating}

\begin{eqnarray} \label{ecom31}
\Sigma \dot{t} &=& \frac{r^2+a^2}{\Delta}[E(r^2+a^2)-aL]-a(aE \sin^2\theta-L),\\
\Sigma \dot{r} &=& \sqrt{R},\label{rad1}\\
\Sigma \dot{\theta} &=& \sqrt{\Theta},\\
\Sigma \dot{\phi} &=& \frac{a}{\Delta}[E(r^2+a^2)-aL]-\left(aE-\frac{L}{\sin^2 \theta}\right),
\end{eqnarray}

\noindent where

\begin{eqnarray} \label{ecom32}
R &=& [E(r^2+a^2)-aL]^2-\Delta[(aE-L)^2+m^2 r^2+ W],\label{rad2}\\
\Theta &=& W \left[\frac{L^2}{\sin^2 \theta}+a^2(m^2-E^2) \right]\cos^2 \theta,
\end{eqnarray}

\noindent and the overdot represents the derivative with respect to the proper time, $W = K-(L-aE)$, with $K$ being an additional constant of movement.

\subsection{Equatorial circular orbits}

Now, let us consider the equatorial plane of the rotating black hole, which is determined by $\theta = \pi/2$. In this region, $\dot{\theta} = 0$, and, thus, $W = 0$, with the radial coordinate being given by

\begin{equation}
r^2 \dot{r} = \pm \sqrt{R},
\end{equation}

\noindent where

\begin{equation}
R = [E(r^2+a^2)-aL]^2-\Delta[(aE-L)^2+m^2 r^2].
\end{equation}

The circular configuration occurs if the radial velocity and acceleration are null. As a consequence, we get, simultaneously, $R(r) = 0$ and $dR/dr = 0$. The energy per unit mass and the axial angular momentum per unit mass are given, respectively, by

\begin{eqnarray}
\frac{E^2_\pm}{m^2} &=& \frac{8 \Delta (a^2- \Delta)^2+ 2 r \Delta \Delta' (a^2- \Delta) - a^2r^2 \Delta'^2 }{r^2[16 \Delta (a^2- \Delta)+ r \Delta'(r \Delta' - 8 \Delta)]} \nonumber \\
&&\frac{\pm 2a \sqrt{2} \Delta \sqrt{(2 a^2-2\Delta + r \Delta')^3}}{r^2[16 \Delta (a^2- \Delta)+ r \Delta'(r \Delta' - 8 \Delta)]},
\end{eqnarray}

\begin{eqnarray}
\frac{L^2_\pm}{m^2} &=& \frac{2 a^2 \Delta^3 - r^2 (r^2+ a^2)^2 \Delta'^2- 2 \Delta^2 \left[ 8 a^2 (r^2+ a^2) + 4 r^4 + a^2 r \Delta' \right]}{r^2[16 \Delta (a^2- \Delta)+ r \Delta'(r \Delta' - 8 \Delta)]} \nonumber \\
&& \frac{2 (r^2+ a^2) \Delta [2 a^2 (r^2+ a^2+ r (3r^2+ a^2)\Delta')]}{r^2[16 \Delta (a^2- \Delta)+ r \Delta'(r \Delta' - 8 \Delta)]} \nonumber \\
&& \mp \frac{2 a \sqrt{2}\Delta \sqrt{(2 a^2-2\Delta + r \Delta')^3} }{r^2[16 \Delta (a^2- \Delta)+ r \Delta'(r \Delta' - 8 \Delta)]} \nonumber \\
&&\left[ (r^2+ a^2) (2(r^2+ a^2)+ r \Delta') - 2(2r^2+ a^2) \Delta \right],
\end{eqnarray}

The squared energy $E^2$ is real if

\begin{equation}
2 a^2- 2 \Delta + r \Delta' \geq 0,
\end{equation}

\noindent which is valid for any $r > 0$. This means that the global monopole parameter does not determine any limit to the existence of circular orbits.

\section{Concluding remarks}
\label{sec5}

We have obtained the solutions corresponding to a charged and static black hole surrounded by a fluid of strings radially pointing. In fact, this fluid is not so realistic, but in an appropriate limit when $\alpha \gg 1$, the obtained solution produces a $1/r$ correction to Newton's gravitational law which can be used, in principle, to explain the flat rotation curves of the galaxies. The solution we obtained we are calling Reissner-Nordstr\"om- de Sitter (anti-de Sitter) black hole with a fluid of strings. The solution depends on the parameter associated with the fluid and is separated into two different analytic forms, namely, for $\alpha = 2$ and $\alpha \neq 2$. We also obtained the solution corresponding to the Kerr-Newman black hole surrounded by a fluid of strings and explicitly gave the two different analytic forms for $\alpha = 2$ and $\alpha \neq 2$.

For the first solution, we analyzed the metric coefficient $f(r)$, whose behaviors are shown in Figs. \ref{Fig1} and \ref{Fig2}, from wich we can see that for $\alpha < 2$, $f(r)$ increases with the radial distance for $\epsilon = + 1$, while it decreases for $\epsilon = - 1$. In the case $\alpha = 2$, $f(r)$ decreases until a certain value of $r$, namely $r = 0.2$ for the parameters choosen, and then it increases for $\epsilon = +1$ and for $\epsilon = -1$, it always decreases. In Fig. \ref{Fig2}, the behaviors of $f(r)$ close to the value $\alpha = 2$ and for $\epsilon = \pm 1$ are shown in the left panel. The behaviors of $f(r)$ are similar to the case shown in the left panel of Fig. \ref{Fig1}. The same happens in which concerns to the right panel.

The effective potential of the geodesic motion also depends on $\alpha$, as expected, as shown in Fig. \ref{Fig3}. As a consequence, the velocity of a particle in an orbit around the black hole depends on $\alpha$, as shown in Fig. \ref{Fig4}.

As to the Kerr-Newman black hole wit fluid of strings, we can observe that the behavior of $\Delta(r)$ depends on the value of $\alpha$, and qualitatively there is also an influence related to the signal of the energy as we can see in Fig \ref{Fig5}, in both panels. We also showed that, depending on the parameter $\alpha$, according to Fig. \ref{Fig6}, the black hole horizons and the static surfaces approaches or not each other. Thus, depending on the values of $\alpha$, the ergoregion has different areas. For $\alpha \leq 2$, the horizons and the static surfaces are more close in the case which $\alpha > 2$, as shown in Fig. 6. In which concerns the equatorial circular orbits, we can have stable or unstable circular orbits depending on the possible value of $\alpha$, which determines the qualitative characteristic of the fluid of strings.

\begin{acknowledgements}
V. B. Bezerra is partially supported by Conselho Nacional de Desenvolvimento Cient\'ifico e Tecnol\'ogico (CNPq) through the research Project nr. 305835/2016-5.
\end{acknowledgements}

\bibliography{refs}

\end{document}